\documentclass[a4paper,12pt,reqno]{amsart}

\usepackage{mathptmx}       
\usepackage{helvet}         
\usepackage{courier}        
\usepackage{hyperref}
\usepackage{amsmath,amssymb}

\def\zeu{\zeta_1}
\def\eu{e_1}
\def\ru{r_1}

\def\N{\mathbf{N}}

\def\R{\mathbf{R}}
\def\C{\mathbf{C}}
\def\x{\textbf{x}}

\def\k{\textbf{k}}

\def\xp{\textbf{x}_\parallel}
\def\yp{\textbf{y}_\parallel}

\def\kp{\textbf{k}_\parallel}

\def\oo{{\bf 0}}
\def\al{\alpha}
\def\del{\delta}
\def\eps{\varepsilon}
\def\vfi{\varphi}
\def\tap{\tau_\pi}

\def\la{\langle}
\def\ra{\rangle}

\def\Gz{G_z}
\def\Gzh{\breve{G}_z}

\def\GKz{\mathcal{G}_z}

\def\A0{A_0}
\def\Aa{A_\al}

\def\Aa{A_\al}
\def\RR0{R_0}
\def\RRa{R_\al}

\def\RRa{R_\al}

\def\RK0{\mathcal{R}_0}
\def\RRK{\mathcal{R}}
\def\Wa{W_\al}
\def\WaK{\mathcal{W}_\al}
\def\Lap{\Delta}
\def\Dom{\mbox{Dom}}
\def\dd{\displaystyle}

\newcommand{\Tr}{\operatorname{Tr}\,}
\def\Fou{\mathcal{F}}

\def\SS{\mathcal{S}}

\def\D0{\mathcal{D}}
\def\Q{\mathcal{Q}}
\def\gal{\tilde{\al}}

\def\dn{g}
\def\en{p}

\def\l{\left}
\def\r{\right}
\def\supp{\mbox{supp}}
\def\vu{\lambda}

\def\Ecf{\omega}
\def\Ec{\mathcal{E}}
\def\CC{\mathcal C}

\renewcommand{\Im}{\operatorname{Im}\,}
\renewcommand{\Re}{\operatorname{Re}\,}
\newcommand{\CO}{\mathbb{C}}
\newcommand{\ve}{\varepsilon}

\title[Relative-Zeta and Casimir energy for a   hyperplane selecting transverse modes]{Relative-Zeta and Casimir energy for a   semitransparent hyperplane selecting transverse modes}

\author{Claudio Cacciapuoti}
\address{DiSAT, Sezione di Matematica, Universit\`a dell'Insubria, via Valleggio 11, I-22100
Como, Italy}
\email{claudio.cacciapuoti@unisubria.it}

\author{Davide Fermi}
\address{Dipartimento di Matematica, Universit\`a di Milano, Via Cesare Saldini 50, I-20133 Milano, Italy }
\email{davide.fermi@unimi.it}

\author{Andrea Posilicano}
\address{DiSAT, Sezione di Matematica, Universit\`a dell'Insubria, via Valleggio 11, I-22100
Como, Italy}
\email{andrea.posilicano@unisubria.it}

\begin{document}

\begin{abstract}
We study  the relative zeta function for the couple of operators $A_0$ and $A_\alpha$, where $A_0$ is the free unconstrained Laplacian in $L^2(\R^d)$ ($d \geq 2$) and $A_\alpha$ is the singular  perturbation of $A_0$ associated to the presence of a delta interaction supported by a hyperplane. In our setting the operatorial parameter $\alpha$, which is related to the strength of the perturbation, is of the kind $\alpha=\alpha(-\Delta_{\parallel})$, where $-\Delta_{\parallel}$ is the free Laplacian in $L^2(\R^{d-1})$. Thus $\alpha$ may depend on the components of the wave vector parallel to hyperplane; in this sense $A_\alpha$ describes a semitransparent hyperplane selecting transverse modes. \\
As an application we give an expression for  the associated thermal Casimir energy. Whenever $\alpha=\chi_{I}(-\Delta_{\parallel})$, where $\chi_{I}$ is the characteristic function of an interval $I$, the thermal Casimir energy can be explicitly computed. 
\end{abstract}

\maketitle

\begin{footnotesize}
 
 \end{footnotesize}

\section{Introduction}
Analytic continuation techniques are well known to be useful   to give a meaning to otherwise divergent series. The most classical example is the Riemann zeta function. The series 
\begin{equation}
\label{rzeta}
\zeta^R(s) := \sum_{n=1}^\infty \frac1{n^s}
\end{equation}
converges only for $s \in \C$ with $\Re s>1$; however, it is well known that the function $s \mapsto \zeta^R(s)$ can be analytically continued to all complex $s\neq 1$. In this way one can formally evaluate the series in Eq. \eqref{rzeta} also for $\Re s < 1$. \\
The same regularization procedure can be used to give a  meaning to divergent series arising when computing   traces of powers of operators, such as $\Tr A^{-s}$. Indeed, when $A$ is a positive, elliptic differential operator with pure point spectrum, which we denote by $\{\lambda_n\}_{n=1}^{\infty}$ ($\lambda_n > 0$ and each eigenvalue is counted with its multiplicity), one can set, in analogy with Eq. \eqref{rzeta}, 
\begin{equation}
\label{zetaA}
	\zeta(A;s) := \Tr A^{-s} = \sum_{n=1}^\infty \frac1{\lambda_n^s}\;.
\end{equation}
The striking feature of the function $\zeta(A;s) $ is that, even though the  series on the r.h.s. converges only for large enough $\Re s$, under certain assumptions on the operator $A$ it can be extended to a meromorphic function with possible poles only on the real line, see \cite{seeley67}. 

When the essential spectrum of the operator $A$ is not empty the regularization procedure described above cannot be applied. This is  the case,  for example, when the operator $A$ is the Laplacian on a non compact manifold; in such a situation the trace $\Tr A^{-s}$ cannot be defined for any $s\in\CO$. \\
Zeta-regularization techniques, however, turn out to be a powerful tool also in these circumstances if  one is interested in the comparison between two operators: an  operator $A$ associated to the ``interacting'' dynamics and a reference  ``non interacting'' or ``free'' operator $A_0$. Both $A$ and $A_0$ are assumed to be nonnegative, they  may have non empty essential spectrum and the traces $\Tr A^{-s}$ and $\Tr A_0^{-s}$ may not be defined. Nevertheless, what may be defined is the \emph{relative zeta function} $\zeta(A,A_0;s) := \Tr (A^{-s}- A_0^{-s})$.\\ 
In certain situations the relative zeta function can be equivalently expressed in terms of the heat semigroups. We recall the following result from   \cite{spreafico-zerbini09}. If the operator  $(A-z)^{-1}- (A_0-z)^{-1}$, with $z$ in the resolvent set of $A$ and $A_0$, is trace class, and such trace has certain asymptotic expansions for $z\to 0$ and $z\to \infty$ (see \cite{spreafico-zerbini09} and  \cite{Mull} for the details), then the formula 
\begin{equation}\label{louisiana}
\zeta(A,A_0;s) = \frac{1}{\Gamma(s)} \int_0^\infty t^{s-1}\Tr (e^{-t A } - e^{-tA_0})dt 
\end{equation}
holds true for $s_0 \leq \Re s \leq s_1$, with $s_0$ and $s_1$ depending on the asymptotic expansions of the trace, and where $\Gamma(s)$ is the Gamma function. \\

The subject of our paper is the study of the relative zeta function for the couple of operators 
 $A_0$ and $A_\alpha$ 
 defined as follows (see Sec. \ref{s:2} for the rigorous definitions):
\begin{itemize}
\item $A_0$ is the free unconstrained Laplacian in $L^2(\R^d)$ ($d \geq 2$).
\item $A_\alpha$ is the Laplacian in the presence of a semitransparent hyperplane selecting transverse modes. More precisely, let  $\pi$ denote  the hyperplane
\begin{equation}
\label{pi}
\pi := \{\x \in \R^d \,|\, x^1 = 0 \}\;;
\end{equation}
this is naturally identified with $\R^{d-1}$. $A_\alpha$ is the self-adjoint operator in $L^2(\R^d)$ which formally corresponds to the Laplacian plus a singular potential supported by the hyperplane $\pi$. In our setting the parameter $\alpha$, which is related to the ``strength'' of the potential, may depend on the components of the wave vector parallel to $\pi$. More precisely, we set  $\alpha = \alpha(-\Delta_\parallel)$, where $-\Delta_\parallel$ is the free unconstrained Laplacian in $L^2(\R^{d-1})$. Heuristically speaking, this indicates that the singular potential supported on $\pi$ acts differently, depending on the transverse  modes (parallel to the hyperplane). Denoting by $\del_1$ the 1-dimensional Dirac delta in $x^1 = 0$, the  operator $\Aa$ formally corresponds to the heuristic expression ``$\Aa = -\Lap + \la \del_1,\cdot\ra\,\del_1 \otimes \al(-\Lap_\parallel)$'' on $L^2(\R^d) \equiv L^2(\R) \otimes L^2(\R^{d-1})$. 
\end{itemize}
We remark that both operators, $A_\alpha$ and $A_0$,  enjoy the translation invariance in the directions $\xp$ parallel to $\pi$. This symmetry of the system has two important consequences. On one side 
the operator $A_\alpha^{-s} - A_0^{-s} $ is not trace class no matter how large $\Re s $ is (a similar remark holds true for the operator $e^{-t A_\alpha } - e^{-tA_0}$). On the other hand it is clear that any relevant (possibly infinite) physical quantity  cannot depend on the coordinate $\xp$ and can be associated to a finite density by averaging on any finite subset of $\pi$. With this remark in mind, and by Eq. \eqref{louisiana}, we infer that the quantity of interest in our analysis is the relative zeta function 
\begin{equation}\label{relzeta1}
\zeu(s) \equiv \zeu(A_\alpha,A_0;s) := \frac{1}{\Gamma(s)} \int_0^{+\infty}\! dt\; t^{s-1} \int_{\R} dx^1 \Q^{rel}(t;x^1,x^1,\oo)\;,
\end{equation}
where 
$\Q^{rel}(t;x^1,y^1,\xp-\yp)$ is the integral kernel of the operator $Q^{rel}(t ):= e^{-tA_\alpha} - e^{-tA_0}$. We remark that here we have used the translation invariance of the system to conclude that the integral kernel $\Q^{rel}$ is a function of $\xp-\yp$.\\
We also note that the integrand function in Eq. \eqref{relzeta1} has been obtained by taking $x^1 = y^1$ and $\xp=\yp$ in the integral kernel $\Q^{rel}(t;x^1,y^1,\xp-\yp)$. Since the integrand function does not depend on $\xp$,  the integral $\int_{\R} dx^1$ could be rewritten in terms of  the average  $|\Omega|^{-1}\int_{\R} dx^1\int_{\Omega} d\xp$, where $\Omega$ is any finite region of $\pi$ of volume $|\Omega|$. In this way, taking the limit $\Omega \to \R^{d-1}$, would reconstruct an ``averaged trace'' of the operator $Q^{rel}(t )$. In the applications, for example when computing the thermal Casimir energy (see Sec. \ref{s:SecCas}), the ``average  argument'' could be made rigorous. A possible approach would be to constrain  the system to a rectangular box of size $L$ along the directions $\xp$, take the average with respect to the volume of the box, and then take the limit $L\to\infty$. Here we do not pursue this  goal, we just recall that the problem of the reduction to a density was already present in the original paper by Casimir \cite{casimir48} as well as in more recent papers such as \cite{Khus}. In the latter work this problem is approached by adding a mass parameter that afterwords is sent to infinity.  \\
Our main result is summed up in Eq.s \eqref{zetaintAC1_1} - \eqref{zetaintAC1_3}, where we give the analytic continuation of the map $s\to \zeu(s)$.\\
As an application, in Sec. \ref{s:SecCas}, we compute the thermal Casimir energy per unit surface for a massless scalar field at temperature $T = 2\pi / \beta$, and discuss an explicit choice of the function $\alpha$.\\ 

In the remaining part of the introduction we discuss the physical motivations of our analysis and several related works. 

Major  applications of the zeta-regularization approach are related to the problem of \emph{zero point oscillations} or \emph{Casimir effect} in Quantum Field Theory (QFT). \\ 
In his 1948 paper \cite{casimir48} the Dutch physicists H. B. G. Casimir pointed out that two parallel, neutral, perfectly conducting plates will show an attractive force. This phenomenon, which was later on named Casimir effect, originates  from the variation of the electromagnetic \emph{zero point energy} due to the boundaries represented by the plates. \\ 
In his setting, Casimir considered a box-shaped cavity with a plate inside, placed parallel to the walls of the cavity. Casimir showed that the plate interacts with the walls through a force (later referred to as \emph{Casimir force}) which is inversely proportional to the cube of the distance between the plate and the walls.  \\ 
The crucial observation in the paper \cite{casimir48} is the following. The energy of the cavity is given by $\frac12 \sum \hbar \omega$ (resp. $\frac12 \sum \hbar \omega'$) where $\omega$ (resp. $\omega'$) are the resonant frequencies of the cavity with (resp. without) the plate inserted in  it, and the sum runs over all the possible frequencies. Even though the sums $\frac12 \sum \hbar \omega$ and $\frac12 \sum \hbar \omega'$ diverge, a finite  value  (which depends on the position of the plate) can be assigned to  the difference of these energies. The Casimir force was indeed computed by taking the derivative of this finite energy difference with respect to the parameter associated to the position of the plate. \\
Nowadays the term Casimir effect refers to a wide class of phenomena that are associated to the variation of the zero point energy or  zero point oscillations in QFT, where a quantized field can be described as a set of oscillators. In a bounded region of the space, for example, the zero point energy of the field is given by a sum of the form $\sum_{j}\omega_j$, where $\omega_j$ are all the possible frequencies of the oscillators and the sum runs over an infinite set of quantum numbers (here denoted by $j$). This series is, in general, divergent. Casimir's approach allows to regularize, by subtraction,  this divergent quantity and extract the relevant information from the regularized energy. \\
The applications of Casimir's regularization are extremely numerous and the literature on the subject is massive. We refer to the monographs \cite{bordag99,bordag-etal09,milton01} for an exhaustive  discussion on this topic and a list of related references. Here we just point out the evident relation between the divergent series in the Casimir effect and zeta-regularization techniques. Indeed, Casimir's force can be computed by regularizing a series of the form given in Eq. \eqref{zetaA} through analytic continuation, and then taking the limit $\lim_{s\to-1}\zeta(A;s)$ (see, e.g., \cite{Eliz}). \\ 
The first attempts to regularize sums involving the eigenvalues of elliptic operators  through analytic continuation  date back to  the works of Minakshisundaram and Pleijel \cite{minakshisundaram49,minakshisundaram-pleijel49}.
  A first example of an application of zeta-regularization  to investigate geometrical properties of manifolds is in \cite{ray-singer71}, where the authors used it to compute the analytic torsion of a smooth, compact manifold. \\
One of the first applications to QFT is in \cite{dowker-critchley76} to compute the effective Lagrangian and the energy-momentum tensor associated to a scalar field in a De Sitter background. In \cite{dowker-critchley76}, the authors point out that this regularization procedure may produce a result different from the one obtained by dimensional regularization. Indeed, shortly afterwards zeta-regularization was discussed by Hawking, see \cite{hawking77}, as a method to resolve the ambiguity in the dimensional regularization of path integrals for fields in curved spacetime. A slightly different (and to some extent  more rigorous) formulation of Hawking's approach was developed by Wald in \cite{wald79}.\\ 
Temperature effects in the classical  Casimir effect were first investigated by Fierz \cite{fierz60} and Mehra \cite{mehra67}. The general dependence on temperature in QFT, instead, was first discussed in \cite{DowKen}. A more recent work in this direction is \cite{OrtSpre}.\\ 
More recently, the zeta-regularization approach was presented in \cite{FP,FPI,FPII,FPIII,FPIV} as a tool to cure the divergences in the vacuum expectation value of both local and global observables in QFT.\\ 
 One of the first attempts at using models with singular potentials (delta-interactions) to compute the energy momentum tensor is in \cite{MaTru}. The same model was taken up again in \cite{Bord}. Delta type interactions intuitively model semitransparent walls. From a mathematical point of view they offer a two-fold  advantage: in a certain sense they are less singular than pure Dirichlet conditions; moreover they produce highly solvable models, i.e. are simple enough to perform explicit computations. In \cite{Graham} the authors compute the Casimir energy of a boson field in the presence of two semitransparent walls in spatial dimension $d=1$ and of a delta interaction supported by a circle in dimension $d=2$. In a similar setting, but in $d=1$ and $d=3$ space dimensions, the Casimir energy and the  pressure   for a massless scalar field are explicitly computed in \cite{Milt}. See \cite{DelCil} for a similar analysis in the case of a delta interaction supported on a cylindrical shell. A systematic analysis of the configuration with two semitransparent walls (with a discussion of the limit in which the boundary conditions become of Dirichlet type) is in \cite{Cast}. \\
We remark that none of the works mentioned in the discussion above use the relative-zeta function regularization scheme. The general theory of the relative-zeta approach was developed by M\"uller in the seminal paper \cite{Mull}.\\ 
When computing the relative zeta function $\zeu(s)$, however, we will not use directly the results in \cite{Mull} but we will follow the equivalent approach presented in  \cite{spreafico-zerbini09}. Our choice relies on the fact that in \cite{Mull}  the relative-zeta function is computed  by exploiting its relation with the difference of the semigroups $e^{-tA_\alpha} - e^{-tA_0}$; in \cite{spreafico-zerbini09}, instead, it is obtained by working with the difference of the resolvents
\[R_\alpha(z) - R_0(z) := (A_\alpha-z)^{-1} - (A_0-z)^{-1} .\]
In our setting the theory of self-adjoint extensions of symmetric operators, see, e.g. \cite{posilicano-jfa}, allows us to obtain an explicit formula for $R_\alpha(z) - R_0(z)$, see Sec. \ref{s:2}, and perform exact calculations in a relatively easy way.\\ 
We conclude by mentioning few works in which  the relative zeta function is used in a setting with singular interactions supported by points: \cite{AlbHalf} where the case of a point potential in the half-space is discussed; and \cite{DelCou} where the authors analyze the combined effect of the Coulomb together with a point potential, both centered at the origin. \\

The paper is structured as follows. In Section \ref{s:2} we introduce the model and obtain an explicit formula for the resolvent of the operator $A_\alpha$, in terms of the resolvent of $A_0$. In Section \ref{s:3}  we obtain a formula for the relative zeta function and study its analytic continuation. In Section \ref{s:SecCas} we give a formula for the thermal Casimir energy, moreover we compute it explicitly in the case in which the function $\alpha$ is the characteristic function of an interval.  We conclude the paper with an Appendix in which we discuss the case $\alpha=constant$.  

\section{The general framework}
\label{s:2}
We work in $d \geq 2$ spatial dimensions and write $\x \equiv (x^i)_{i=1,...,d}$ to denote  points in $\R^d$. We identify the points of the plane $\pi$ defined in Eq. \eqref{pi} with $\xp \equiv (x^2,...,x^d) \in \R^{d-1}$. Moreover, we shall use the notations
\begin{equation}\label{notation_x}
(x^1,\xp) \equiv \x \,\in\, \R^d \equiv \R \times \R^{d-1}.
\end{equation}
We denote by $\Fou$ and $\Fou^{-1}$ the distributional Fourier and inverse Fourier transform defined on integrable functions as 
\[\Fou \vfi(\k) := \int_{\R^d} d\x\;e^{- i \k \cdot \x} \vfi(\x)\;, \qquad \Fou^{-1} \vfi(\x) := \int_{\R^d} d\k\;{e^{i \k \cdot \x} \over (2\pi)^{d}}\;\vfi(\k).\]
Notice that, with the above choice, the convolution of two functions $\vfi,\psi$ fulfills
\[ \Fou(\vfi \ast \psi)(\k) =  \Fou \vfi(\k) \, \Fou \psi(\k)\;.\]
The free Laplacian on $\R^d$ is the self-adjoint operator $$\A0 := -\Lap : \Dom(\A0) \subset L^2(\R^d) \to L^2(\R^d)\,,$$ where $\Dom(\A0) = H^2(\R^d)$ (the Sobolev space of order two); the associated resolvent is the bounded operator $$\RR0(z) := (\A0 - z)^{-1} : L^2(\R^d) \to H^2(\R^d)\,,\qquad z \in \C \setminus [0,+\infty)\,.
$$ \\
Throughout the paper we consider the natural determination of the argument for complex numbers, i.e. $\arg : \C \setminus [0,+\infty) \to (0,2\pi)$; furthermore, for any $z \in \C \setminus [0,+\infty)$, we always use the notation $\sqrt{z}$ to denote the principal square root, i.e. the one with positive imaginary part. \\
As well known, the action of $\RR0(z)$ can be expressed in terms of the corresponding convolution kernel as $
\RR0(z) \vfi = \RK0(z) \ast \vfi$ where $\RK0(z;\x) = e^{i\sqrt z |\x|}/(4\pi |\x|)$. We also recall that the Fourier transform of $\RK0(z;\cdot)$ is given by
$(\Fou \RK0(z))(\k) = (|\k|^2 - z)^{-1}$.
\\
Together with the notation introduced in \eqref{notation_x},  we shall often write
\begin{equation*}
(k_1,\kp) \equiv \k \,\in\, \R^d \equiv \R \times \R^{d-1}\;.
\end{equation*}
Next, let us consider a non-negative, piecewise continuous, and compactly supported function $\alpha$ such that, for some $\delta>0$, it holds true:  
\begin{equation} \label{al1}
\al(\rho) > \delta \quad \forall \rho \in \supp(\alpha). 
\end{equation}
The trace on the hyperplane $\pi=\{\x \in \R^d \,|\, x^1 = 0 \}$ is the unique linear bounded operator $$\tap : H^r(\R^d) \to H^{r-{1 \over 2}}(\R^{d-1})\,,\qquad r > 1/2\,,
$$ 
such that, for any continuous $\vfi$ there holds
\begin{equation*}
(\tap \vfi)(\xp) = \vfi(0,\xp)\;, \qquad \xp \in \R^{d-1}\;.
\end{equation*}
Here and below $H^{s}(\Omega)$, $s\in\R$, denotes the usual scale of Sobolev-Hilbert spaces on the open subset $\Omega\subseteq\R^{n}$. In particular, considering the Sobolev spaces on the half spaces $\R^d_\pm := \{\x \in \R^d\,|\,\pm x^1 > 0 \}$, one introduces the lateral traces $$\tap^{\pm} : H^r(\R^d_{\pm}) \to H^{r-{1 \over 2}}(\R^{d-1})\,,\qquad r > 1/2\,,$$ defined as the unique linear bounded operators such that, for any continuous (up to the boundary) function on $\R^{d}_{\pm}$ there holds  
\begin{equation*}
(\tap^{\pm} \vfi)(\xp) = \vfi(0_{\pm},\xp)\;, \qquad \xp \in \R^{d-1}\;.
\end{equation*}
Setting 
$$H^r(\R^d\!\setminus\!\pi):=H^r(\R^d_-) \oplus H^r(\R^d_+)\,,$$ one has that  $\vfi=\vfi_{-}\oplus\vfi_{+}\in H^{r}(\R^d\!\setminus\!\pi)$, $1/2<r<3/2$, belongs to $H^{r}(\R^d)$ if and only if $\tap^{-}\vfi_{-}=\tap^{+}\vfi_{+}$; in this case $\tap^{\pm}\vfi_{\pm}=\tap\vfi$.  
\\
To proceed, consider the free Laplacian on $\R^{d-1}$, indicated hereafter with $-\Lap_\parallel$; since its spectrum coincides with the half-line $[0,+\infty)$, one can define via standard functional calculus the bounded self-adjoint operator 
$$
\al(-\Lap_\parallel) : L^2(\R^{d-1}) \to L^2(\R^{d-1})\,.
$$ 
We use such an operator to define a self-adjoint singular perturbation of the free Laplacian 
$\Aa:\Dom(\Aa) \subset L^2(\R^d) \to L^2(\R^d)$ with domain
\begin{equation}
\begin{split}
\Dom(\Aa)= &\{\vfi=\vfi_{-}\oplus\vfi_{+}\!\in\!H^2(\R^d\!\setminus\!\pi) \;|\;\tap^{-}\vfi_{-}=\tap^{+}\vfi_{+}\\
&\qquad\tap^{+}\partial_{1}\vfi_{+}-\tap^{-}\partial_{1}\vfi_{-}=\al(-\Lap_{\parallel}) \tap\vfi\} \,.
\end{split}
\label{AaDef}
\end{equation}
Similar models, with momentum dependent delta-interactions supported on spherical shells, were studied in \cite{Sha1,Sha2,Sha3}. 
\\
Let us consider, for any $z \in \C \!\setminus\! [0,+\infty)$, the bounded operator
\begin{equation*}
\Gzh : L^2(\R^d) \to H^{3/2}(\R^{d-1})\;, \qquad \Gzh \vfi := \tap \RR0(z) \vfi.
\end{equation*}
Next, consider the adjoint of $\breve{G}_{\bar z}$ with respect to the $H^{-3/2}(\R^{d-1})$-$H^{3/2}(\R^{d-1})$ duality $(\cdot|\cdot)$, that is the unique bounded operator 
\begin{equation*}
\Gz : H^{-3/2}(\R^{d-1}) \to L^2(\R^d)\,, 
\end{equation*}
fulfilling
\begin{equation*}
\la \Gz q | \vfi \ra_{L^2(\R^d)} = ( q | \breve{G}_{\bar z}\vfi) \qquad \mbox{ $q \in H^{-3/2}(\R^{d-1})$, $\vfi \in L^2(\R^d)$}.
\end{equation*}
One can easily check that $G_{z}$ corresponds to the single layer operator of the hyperplane $\pi$. 
\\
Let us now introduce a convenient representation of $R_{0}(z)$; since it is a bounded operator, it suffices to consider its action on any $\vfi=\vfi_{1}\otimes\vfi_{\parallel}$ belonging to the dense subset  $\SS(\R)\otimes\SS(\R^{d-1})$. Recalling the explicit representation of the kernel of the resolvent of the free $1$-dimensional Laplacian $\Delta_{1}$, one has 
\begin{align*}
&(R_{0}(z)\varphi_{1}\otimes\varphi_{\parallel})\,(x^{1},\xp)=((A_{0}-z)^{-1}\varphi_{1}\otimes\varphi_{\parallel})\,(x^{1},\xp)\\
=&\frac{1}{(2\pi)^{d}}\,\int_{\R^{d-1}}d\kp\;e^{i\,\kp\cdot\yp}\Fou\vfi_{\parallel}(\kp)\!\!\left(\int_{\R}dk_1\;
\frac{ e^{ik_{1}x^{1}}\Fou\varphi_{1} (k_{1})}{k_{1}^2+|\kp|^2-z}\right)\\
=&\frac{1}{(2\pi)^{d-1}}\,\int_{\R^{d-1}}d\kp\;e^{i\,\kp\cdot\yp}\Fou\vfi_{\parallel}(\kp)
\left(\left(-\Delta_{1}+|\kp|^2-z\right)^{-1}\vfi_{1}\right)(x^{1})\\
=&\frac{1}{(2\pi)^{d-1}}\,\int_{\R^{d-1}}d\kp\;e^{i\,\kp\cdot\yp}\Fou\vfi_{\parallel}(\kp)
\left(\int_{\R}dy^1\;\frac{ i e^{i\sqrt{z-|\kp|^2}\ |x^{1}-y^{1}|}}{2\sqrt{z-|\kp|^2}}\ \vfi_{1}(y^{1})\right)\\
=&\frac{i}{2}\,\left(\left(\int_{\R}dy^1\;\vfi_{1}(y^{1})\,
e^{i|x^{1}-y^{1}|\,(z+\Delta_{\parallel})^{1/2}}\right)\,(z+\Delta_{\parallel})^{-1/2}
\vfi_{\parallel}\right)(\xp)\, .
\end{align*}
This gives
\begin{equation*}
\label{Gz0}
(\breve{G}_{z}\vfi_{1}\otimes\vfi_{\parallel})(\xp)=
\frac{i}{2}\,\left(\left(\int_{\R}dy^1\;\varphi_{1}(y^{1})\,
e^{i|y^{1}|\,(z+\Delta_{\parallel})^{1/2}}\right)\,(z+\Delta_{\parallel})^{-1/2}
\vfi_{\parallel}\right)(\xp)
\end{equation*}
and so
\begin{equation}\label{Gz}
({G}_{z}q)(x^{1},\xp)=\frac{i}{2}\,\left(e^{i|x^{1}|\,(z+\Delta_{\parallel})^{1/2}}(z+\Delta_{\parallel})^{-1/2}q\right)(\xp)\,.
\end{equation}
We note that the corresponding integral kernels are of convolution type in the variables $\xp$ and $\yp$, i.e. 
\[(\breve{G}_{z}\vfi)(\xp)=\int_{\R} dy^1\int_{\R^{d-1}}\!\! d\yp\; \GKz(y^1,\xp-\yp) \vfi(y^1,\yp)=\int_{\R} dy^1(\GKz(y^1,\cdot) \ast \vfi(y^1,\cdot)) (\xp)\;, \]
\[(G_{z}q)(x^1,\xp)=\int_{\R^{d-1}} d\yp\; \GKz(x^1,\xp-\yp)q(\yp)=(\GKz(x^1,\cdot) \ast q)(\xp)\;. \]
Moreover the Fourier transform (on $\xp$)  of the function  $\GKz(x^1,\xp) $ is given by 
\[(\Fou \GKz(x^1,\cdot ))(\kp) = \frac{ie^{i|x^{1}|\,\sqrt{z-|\kp|^2}}}{2\sqrt{z-|\kp|^2}}.\]
%
%
By Eq. \eqref{Gz}, one infers that $\tap \Gz$, $z \in \C \setminus [0,+\infty)$, extends to a well defined pseudodifferential operator $M_{z}$ of order $(-1)$ defined on the whole scale of Sobolev-Hilbert spaces on $\R^{d-1}$:
$$
M_{z}: H^{r}(\R^{d-1})\to H^{r+1}(\R^{d-1})\,,\quad M_{z}:=\frac{i}{2}\,(z+\Delta_{\parallel})^{-1/2}\,.
$$
Then, by using $M_{z}$, we define, for any  $z \in \C \setminus [0,+\infty)$,  
\begin{equation*}
\label{Walpha}
\Wa(z) := - \,\al(-\Lap_\parallel) \Big(1+\al(-\Lap_\parallel) M_{z}\Big)^{\!\!-1}\; : \; L^2(\R^{d-1}) \;\to\; L^2(\R^{d-1})\,.
\end{equation*}
Notice that $\Wa(z)$ is a bounded operator since, by functional calculus, $\Wa(z)=w_{z}(-\Delta_{\parallel})$, where 
\[
w_{z}(\rho)=-\frac{\alpha(\rho)\sqrt{z-\rho}}{\sqrt{z-\rho}+i\alpha(\rho)/2}
\]
and $w_{z}\in L^{\infty}(0,+\infty)$ for any $z \in \C \setminus [0,+\infty)$. Indeed, the associated convolution kernel is given by $\WaK(z;\xp-\yp)$, with $(\Fou \WaK(z;\cdot))(\kp) = w_{z}(|\kp|^2)$.
\\
Finally, for any $z \in \C\setminus [0,+\infty)$ we define  the bounded linear operator $\RRa(z)$ by 
\begin{equation*}
\label{Ralpha}
\RRa(z) : L^{2}(\R^{d})\to L^{2}(\R^{d})\,,\quad
\RRa(z) = \RR0(z) + \,\Gz\,\Wa(z)\, \Gzh\;.
\end{equation*}
By the results provided in \cite[section 2]{posilicano-jfa}, applied to the case in which the map there denoted by 
$\tau$ is given by $\tau:=\sqrt \alpha(-\Delta_{\parallel})\tau_{\pi}$, one gets that $\RRa$ is a pseudo-resolvent, i.e. it satisfies the resolvent identity $\RRa(w)-\RRa(z)=(z-w)\RRa(w)\RRa(z)$. Moreover, by Eq. \eqref{Gz},  
\begin{equation}\label{jump}
\tap^{-}\partial_{1}G_{z}q-\tap^{+}\partial_{1}G_{z}q=q
\end{equation}
and so Ran$(G_{z})\cap H^{2}(\R^{d})=\{0\}$, i.e. $\RRa(z)$ is injective. In conlcusion, since    
$\RRa(\bar z)=\RRa(z)^{*}$, $\RRa(z)$ is the resolvent of the ($z$-independent) self-adjoint operator $\Aa:=\RRa(z)^{-1}+z$ defined on the domain $\Dom(\Aa):=$Ran$(\RRa(z))$. Setting 
$G:=G_{-1}$ and $\Wa:=\Wa(-1)$, one has 
\begin{equation}\label{dom}
\Dom(\Aa)=\{\vfi\in L^{2}(\R^{d})\;|\;\vfi=\vfi_{0}+G\Wa\tap\vfi_{0}\,,\ \vfi_{0}\in H^{2}(\R^{d})\}
\end{equation}
and, by the identity $\RRa(-1)(A_{0}+1)\vfi_{0}=\vfi_{0}+G\Wa\tap\vfi_{0}$, 
\begin{equation}\label{Aa}
(\Aa+1)\vfi=(A_{0}+1)\vfi_{0}\,.
\end{equation}
The representation of $\Dom(\Aa)$ given in \eqref{dom} coincides with \eqref{AaDef} by 
$\tap^{-}Gq=\tap^{+}Gq$ (which is consequence of Eq. \eqref{Gz}) and by the identities (here we use Eq.  \eqref{jump} and the definition of $\Wa$)
\begin{align*}
&\tap^{+}\partial_{1}\vfi-\tap^{-}\partial_{1}\vfi =\tap^{+}\partial_{1}G\Wa\tap\vfi_{0}-\tap^{-}\partial_{1}G\Wa\tap\vfi_{0} = -\Wa\tap\vfi_{0}\\
= & -(1+\alpha(-\Delta_{\parallel})\tap G)\Wa\tap\vfi_{0}+\alpha(-\Delta_{\parallel})\tap G\Wa\tap\vfi_{0}\\
= & \alpha(-\Delta_{\parallel})\tap \vfi_{0}+\alpha(-\Delta_{\parallel})\tap G\Wa\tap\vfi_{0}=\alpha(-\Delta_{\parallel})\tap \vfi\,.
\end{align*}
Moreover, since $(-\Delta+1) G=\delta_{\pi}$, where $\delta_{\pi}$ denotes the tempered distribution defined by $\delta_{\pi}(\phi):=\int_{\R^{d-1}}\phi(0,\xp)\,d\xp$, $\phi\in\SS(\R^{d})$, by Eq. \eqref{Aa} one has
$$ \Aa\vfi=A_{0}\vfi_{0}+G\Wa\tap\vfi_{0}=-\Delta\vfi-\Wa\tap\vfi_{0}\delta_{\pi}=
-\Delta\vfi+\alpha(-\Delta_{\parallel})\tap\vfi\delta_{\pi}\,. $$

\section{The relative zeta function}
\label{s:3}
In this section we obtain an explicit expression for the relative zeta function $\zeu(s)$. The formula is given in Eq. \eqref{zetaint} and expresses the relative zeta function as an integral of the relative spectral measure $\eu(v)$. This identity holds true for the values of $s$ in the strip \eqref{Res1}. The function $\eu(v)$ is defined in Eq. \eqref{defev} and computed explicitly in Sec. \ref{s:ev}. In the same section we also obtain the asymptotic expansions of the function $\eu(v)$ for $v\to0^+$ and $v\to+ \infty$. We will use these asymptotic expansions in Sec. \ref{SecZeta} to obtain the analytic continuation of the function $\zeu(s)$ to the strip defined by Eq. \eqref{zetaintAC1_2}. 

Denote by $R^{rel}(z)$ the operator 
\[ R^{rel}(z) := R_\alpha(z) - R_0 (z) \qquad z \in \C \setminus [0,+\infty). \]
The integral kernel of $R^{rel}(z)$ is of convolution type on the variables $\xp$ and $\yp$ and it is given by 
\begin{align}
& \RRK^{rel}(z;x^1,y^1,\xp - \yp) = (\GKz(x^1,\cdot)\ast \WaK(z;\cdot)  \ast \GKz(y^1,\cdot ))(\xp - \yp)\nonumber \\ 
=& {1\over 4(2\pi)^{d-1}} \int_{\R^{d-1}}\! d\kp\;  e^{i\kp \cdot (\xp-\yp)} \frac{\al(|\kp|^2)\, e^{i(|x^1|+|y^1|)\sqrt{z-|\kp|^2}}}{\sqrt{z - |\kp|^2} \left( \sqrt{z-|\kp|^2} + i \al(|\kp|^2)/2 \right)} \;. \label{toolong}
\end{align}
To compute the relative zeta function defined in Eq. \eqref{relzeta1} we will compute first  the function 
\begin{equation} \label{r1z}
\ru(z) := \int_\R dx^1\; \RRK^{rel}(z;x^1,x^1,\oo) \qquad z \in \C \setminus [0,+\infty). 
\end{equation}
Then we will show that the \emph{relative spectral measure}\footnote{We remark that here we are slightly abusing terminology, as ``relative spectral measure''  usually denotes  the function $e(v) := {v \over i\pi} \lim_{\eps \to 0^+} \Tr(R(v^2+i \eps) - R_0(v^2 -i \eps))$.}
\begin{equation}\label{defev}
\eu(v) := {v \over i\pi} \lim_{\eps \to 0^+} \Big[\ru(v^2+i \eps) - \ru(v^2 -i \eps)\Big]
\end{equation}
is well defined for any $v>0$, see Sec. \ref{s:ev}. Finally, is Sec. \ref{s:zeta1},  we shall  prove  that the function $\zeu(s)$ can be expressed through the following fundamental formula 
\begin{equation}
\zeu(s) = \int_0^{+\infty} dv\;v^{-2s}\;\eu(v) \label{zetaint}
\end{equation}
for any complex $s$ in the strip 
\begin{equation}
\l\{ s \in \C \,\;\Big|\; - {1 \over 2} < \Re s < {d-1\over 2} \r\}. \label{Res1}
\end{equation}
Formula \eqref{zetaint}, together with the asymptotic expansions of $\eu(v)$ for $v\to 0^+$ and $v\to +\infty$, see Eq.s \eqref{ev0} and \eqref{evinf}, can be used to obtain the analytic continuation of the map $s\to \zeu(s) $ outside the strip \eqref{Res1}, see Eq.s \eqref{zetaintAC1_1} - \eqref{zetaintAC1_3}.  \\

We start by computing function $\ru(z)$ defined in Eq. \eqref{r1z}. \\
For notational convenience, we introduce the rescaled function
\begin{equation}
\gal(\rho) := {1 \over 2}\;\al(\rho) \qquad (\rho \in [0,+\infty)\,)\;. \label{galdef}
\end{equation}
Setting $y^1 = x^1 $ and $\yp = \xp$ in Eq. \eqref{toolong}, and integrating over $x^1$ we are left with   
\begin{equation}\label{rzkp}
\ru(z) = {i \,\pi \over (2\pi)^{d}} \int_{\R^{d-1}}\! d\kp\; {\gal(|\kp|^2) \over (z - |\kp|^2) \l(\sqrt{z-|\kp|^2} + i \gal(|\kp|^2)\r)} \;.
\end{equation}
Note that for any $z \in \C \setminus [0,+\infty)$, we have $\Im \sqrt{z-|\kp|^2} >0$ and recall that the function $\alpha(\rho)$ is compactly supported by assumption. Hence, one can exchange order of integration and perform the integral over $x^1$. \\ 
Next, passing to polar coordinates and considering the change of variable $\rho := |\kp|^2 \in (0,+\infty)$, one obtains\vspace{-0.05cm}
\begin{equation}
\ru(z) = {i \,\pi^2 \over (2\pi)^{{d + 3 \over 2}}\, \Gamma({d-1 \over 2})}\int_{0}^{+\infty} d\rho\; {\rho^{d-3 \over 2}\; \gal(\rho) \over (z - \rho) (\sqrt{z-\rho} + i \gal(\rho))} \;, \label{rz}
\end{equation}
where  $\Gamma(\cdot )$ denotes  the Euler Gamma function.

\subsection{The relative spectral measure and its asymptotic expansion}\label{s:ev}
In this section we obtain an explicit formula for the  function $\eu(v)$ defined in Eq. \eqref{defev} (see Eq. \eqref{evExp} below). Then compute its asymptotic expansion for $v\to 0^+$, see Eq. \eqref{ev0}, and for $v\to +\infty$, see Eq. \eqref{evinf}. \\ 

First of all, let us point out the trivial identity 
\begin{equation*}
{\gal(\rho) \over (z- \rho) (\sqrt{z-\rho} + i\gal(\rho))} = 
-\frac{i}{z-\rho} +\frac{i}{\sqrt{z-\rho}(\sqrt{z-\rho}+ i\gal(\rho))}.\end{equation*}
In view of the above identity, and recalling Eq. \eqref{rz}, the difference $\ru(v^2+i \eps) - \ru(v^2 -i \eps)$ can be expressed via simple algebraic manipulations as
\begin{equation*}
\ru(v^2+i \eps) - \ru(v^2 -i \eps) =  \frac{i \,\pi^2}{(2\pi)^{{d + 3 \over 2}} \Gamma\left(\frac{d-1}2\right)} \big(E_1(\ve;v) + E_{2}(\ve,v) +E_3(\ve;v)\big)
\end{equation*}
where 
\[ E_1(\ve;v) := - \int_{\supp\alpha} d\rho \rho^{\frac{d-3}{2}}  {2\eps \over (v^2 - \rho)^2 + \eps^2}\;, \]
\[ E_2(\ve;v) :=i \int_{\supp\alpha} d\rho   \frac{\rho^{\frac{d-3}{2}}}{\sqrt{v^2-\rho +i\ve}+i\gal(\rho)} 
\left(\frac{1}{\sqrt{v^2-\rho+i\ve}}- \frac{1}{\sqrt{v^2-\rho-i\ve}}\right)\,, \]
\[\begin{aligned}
E_3(\ve;v) :=&i \int_{\supp\alpha} d\rho   \frac{\rho^{\frac{d-3}{2}}}{\sqrt{v^2-\rho -i\ve}} \left(\frac{1}{\sqrt{v^2-\rho+i\ve}+i\gal(\rho)}- \frac{1}{\sqrt{v^2-\rho-i\ve}+i\gal(\rho)}\right).
\end{aligned}\]
We study the convergence for $\ve\to0^+$ term by term.\\
For the term $E_1(\ve;v)$ we use the fact that ${2\eps \over (v^2 - \rho)^2 + \eps^2} \to \pi \delta(\rho-v^2)$ and the fact that, $\rho^{\frac{d-3}{2}} \chi_{\supp\alpha}(\rho)$ is piecewise continuous (here and below $\chi_{\supp\alpha}(\cdot)$ denotes the characteristic function of $\supp\alpha$), to get  (see, e.g., \cite[Ex. 1.13]{kanwal})
\begin{equation}\label{limE1}
\lim_{\ve\to 0^+}E_1(\ve;v) = -2\pi v^{d-3}\chi_{\supp\alpha}(v^2)
\end{equation}
To compute the limit of $E_2(\ve;v)$ and $E_3(\ve;v)$ requires a bit more work. We start by recalling the following inequalities, holding for some $C>0$:
\begin{equation}\label{mambo1}
|\sqrt{\lambda + i\ve}-\sqrt{\lambda - i\ve}   - 2\sqrt\lambda | \leq C \sqrt \ve  \qquad\forall \lambda  > 0\;;
\end{equation}
\begin{equation}\label{mambo2}
|\sqrt{\lambda + i\ve}-\sqrt{\lambda - i\ve}  | \leq C \sqrt \ve  \qquad\forall \lambda  < 0\;.
\end{equation}
Next, we analyze the term  $E_2(\ve;v)$. We note the identity 
\[\begin{aligned}
& E_2(\ve;v) - i \int_{\supp\alpha\cap(0,v^2)} d\rho   \frac{\rho^{\frac{d-3}{2}}}{\sqrt{v^2-\rho +i\ve}+i\gal(\rho)} 
\frac{2\sqrt{v^2-\rho}}{\sqrt{(v^2-\rho)^2+\ve^2}} \\ 
=&\;i \int_{\supp\alpha\cap(0,v^2)} d\rho   \frac{\rho^{\frac{d-3}{2}}}{\sqrt{v^2-\rho +i\ve}+i\gal(\rho)} 
\frac{\sqrt{v^2-\rho +i\ve}-\sqrt{v^2-\rho - i\ve}-2\sqrt{v^2-\rho}}{\sqrt{(v^2-\rho)^2+\ve^2}} \\ 
&+i \int_{\supp\alpha\cap(v^2,+\infty)} d\rho   \frac{\rho^{\frac{d-3}{2}}}{\sqrt{v^2-\rho +i\ve}+i\gal(\rho)} 
\frac{\sqrt{v^2-\rho +i\ve}-\sqrt{v^2-\rho - i\ve}}{\sqrt{(v^2-\rho)^2+\ve^2}},
\end{aligned}\] 
where we used the fact that, due to our definition of the square root, we have that 
\[\sqrt{v^2-\rho+i\ve}\sqrt{v^2-\rho-i\ve} = - \sqrt{(v^2-\rho^2) +\ve^2 }.\]
 To proceed, let us recall that the square root is taken with positive imaginary part and that $\al(\rho) > \del$ for all $\rho \in \supp\al$, see Eq. \eqref{al1}; then, by using the inequalities \eqref{mambo1} - \eqref{mambo2}, we infer
\[\begin{aligned}
& \left|E_2(\ve;v) - i \int_{\supp\alpha\cap(0,v^2)} d\rho \frac{\rho^{\frac{d-3}{2}}}{\sqrt{v^2-\rho +i\ve}+i\gal(\rho)} 
\frac{2\sqrt{v^2-\rho}}{\sqrt{(v^2-\rho)^2+\ve^2}} \right | \\ 
\leq & \;\frac{C}{\delta}  \left(\int_{\supp\alpha} d\rho  \rho^{\frac{d-3}{2}} 
\frac{\sqrt \ve}{\sqrt{(v^2-\rho)^2+\ve^2}}  \right) \leq \frac{C \ve^{1/4}}{\delta}  \int_{\supp\alpha} d\rho  \rho^{\frac{d-3}{2}} 
\frac{1}{|v^2-\rho|^{3/4}} \\ 
\leq & \; C(\alpha, d) \ve^{1/4}\;.
\end{aligned}\] 
Since, by dominated convergence, one has 
\[\begin{aligned}
&\lim_{\ve\to 0^+} i \int_{\supp\alpha\cap (0,v^2)} d\rho   \frac{\rho^{\frac{d-3}{2}}}{\sqrt{v^2-\rho +i\ve}+i\gal(\rho)} 
\frac{2\sqrt{v^2-\rho}}{\sqrt{(v^2-\rho)^2+\ve^2}} \\  
=& \;i \int_{\supp\alpha\cap (0,v^2)} d\rho   \frac{\rho^{\frac{d-3}{2}}}{\sqrt{v^2-\rho}+i\gal(\rho)} 
\frac{2}{\sqrt{v^2-\rho}}\; ,
\end{aligned}\] 
we conclude that 
\begin{equation}\label{limE2}
\lim_{\ve\to 0^+}E_2(\ve,v) = 
 i \int_{\supp\alpha\cap(0,v^2)} d\rho \frac{\rho^{\frac{d-3}{2}}}{\sqrt{v^2-\rho}+i\gal(\rho)} 
\frac{2}{\sqrt{v^2-\rho}}\;.
\end{equation}
Reasoning in a similar way, let us consider the identity 
\[\begin{aligned}
& E_3(\ve;v)  - i \int_{\supp\alpha\cap (0,v^2)} d\rho  
\frac{\rho^{\frac{d-3}{2}} (v^2-\rho -i\ve)^{-1/2} 2\sqrt{v^2-\rho}}{\big(\Re \sqrt{v^2-\rho+i\ve}\big)^2+\big(\Im \sqrt{v^2-\rho+i\ve}+\gal(\rho)\big)^2 } \\ 
=&i \int_{\supp\alpha\cap (0,v^2)} d\rho   \frac{\rho^{\frac{d-3}{2}}(v^2-\rho -i\ve)^{-1/2}(\sqrt{v^2-\rho +i\ve}-\sqrt{v^2-\rho - i\ve}-2\sqrt{v^2-\rho})}{\big(\Re \sqrt{v^2-\rho+i\ve}\big)^2+\big(\Im \sqrt{v^2-\rho+i\ve}+\gal(\rho)\big)^2} \\ 
&+i \int_{\supp\alpha\cap (v^2,+\infty)} d\rho   \frac{\rho^{\frac{d-3}{2}}(v^2-\rho -i\ve)^{-1/2}(\sqrt{v^2-\rho +i\ve}-\sqrt{v^2-\rho - i\ve})}{\big(\Re \sqrt{v^2-\rho+i\ve}\big)^2+\big(\Im \sqrt{v^2-\rho+i\ve}+\gal(\rho)\big)^2}\;,
\end{aligned}\] 
where we used the fact that $\Re \sqrt{v^2-\rho-i\ve} = - \Re \sqrt{v^2-\rho + i\ve}$ and  $\Im \sqrt{v^2-\rho-i\ve} = \Im \sqrt{v^2-\rho + i\ve}$. This allows us to infer
\[\begin{aligned}
& \left| E_3(\ve;v)- i \int_{\supp\alpha\cap (0,v^2)} d\rho  
\frac{\rho^{\frac{d-3}{2}} (v^2-\rho -i\ve)^{-1/2} 2\sqrt{v^2-\rho}}{\big(\Re \sqrt{v^2-\rho+i\ve}\big)^2+\big(\Im \sqrt{v^2-\rho+i\ve}+\gal(\rho)\big)^2 }  \right| \\ 
\leq &\; \frac{\sqrt \ve}{\delta^2} \int_{\supp\alpha} d\rho   \frac{\rho^{\frac{d-3}{2}}}{|\sqrt{v^2-\rho -i\ve}|} 
 \leq \frac{\sqrt \ve}{\delta^2} \int_{\supp\alpha} d\rho   \frac{\rho^{\frac{d-3}{2}}}{\sqrt{|v^2-\rho|}} \leq C(\alpha,d) \sqrt \ve.  
\end{aligned}\] 
As before, by using the  dominated convergence theorem, one can prove that 
\begin{equation}\label{limE3}
\lim_{\ve\to0^+} E_3(\ve;v) = - i \int_{\supp\alpha\cap (0,v^2)} d\rho \frac{\rho^{\frac{d-3}{2}}}{\sqrt{v^2-\rho}} 
\frac{2\sqrt{v^2-\rho}}{v^2-\rho+\gal(\rho)^2 } .
\end{equation}
Summing up the limits \eqref{limE2} and \eqref{limE3}, we conclude 
\[
\lim_{\ve\to0^+} \big(E_2(\ve;v) + E_3(\ve;v)) = \int_0^{v^2} d\rho   \frac{\rho^{\frac{d-3}{2}}}{\sqrt{v^2-\rho}} 
\frac{2\gal(\rho)}{v^2-\rho+\gal(\rho)^2 } .
\]
Taking into account also the limit \eqref{limE1} we get the following expression:
\begin{equation}
\eu(v) = {\pi\,v \over (2\pi)^{{d + 1 \over 2}} \Gamma({d-1 \over 2})} \Big[- v^{d-3}\,\chi_{\supp\al}(v^2) + I_{\al}(v^2)\,\chi_{(0,+\infty)}(v)\Big]\;, \label{evExp} 
\end{equation}
where we introduced the notation
\begin{equation}
I_{\al}(\vu) := {1 \over \pi}\int_{0}^{\vu} d\rho\; {\rho^{d-3 \over 2}\,\gal(\rho) \over \sqrt{\vu - \rho}\; (\vu - \rho + \gal(\rho)^2)} \qquad \mbox{for any $\vu \in (0,+\infty)$}\;. \label{evI}
\end{equation}
Before proceeding, let us stress that due to the assumptions on $\al$, the integral $I_{\al}$ can be easily checked to be finite and positive for any $\vu \in (0,+\infty)$. In fact, by Lebesgue's dominated convergence theorem, one can infer that the map $\vu \mapsto I_{\al}(\vu)$ is continuous on $[0,+\infty)$. \\
Next, let us pass to discuss the asymptotic behavior of the map $\vu \mapsto I_\al(\vu)$ for $\vu \to 0^+$ and $\vu \to +\infty$. For $\vu \to0^+$, we have that 
\[
I_{\al}(\vu) \leq  {\|\alpha\|_{\infty} \over \pi \delta^2}\int_{0}^{\vu} d\rho\; {\rho^{d-3 \over 2}\,\over \sqrt{\vu - \rho}\; }\leq  {\|\alpha\|_{\infty}  \vu^{d-3 \over 2} \over \pi \delta^2}\int_{0}^{\vu} d\rho\; {1 \over \sqrt{\vu - \rho}\; }\leq C(\alpha)  \vu^{d-2 \over 2} .
\]
Hence,
\begin{equation*}
I_{\al}(\vu) = O(\vu^{{d-2 \over 2}}) \qquad \mbox{for $\vu \to 0^+$} \;. \label{evIzer}
\end{equation*}
Next we discuss the asymptotic behavior of $I_\al(\vu)$ in the limit $\vu \to +\infty$; in particular, we shall  show that  that there exits a family of real coefficients $(\en_n)_{n = 0,1,2,...}$ such that, for any $N \in \{0,1,2,...\}$,
\begin{equation}
I_{\al}(\vu) = \vu^{-3/2}\! \sum_{n = 0}^N \en_n\;\vu^{-n} + O(\vu^{-N-{5 \over 2}}) \quad \mbox{for $\vu \to +\infty$}\;; \label{evIinf}
\end{equation}
for example, the first two coefficients are
\begin{equation}
\begin{array}{c}
\dd{\en_0 := {1 \over \pi} \int_{\supp \al}\! d\rho\; \rho^{d-3 \over 2}\;\gal(\rho)\;,} \vspace{0.1cm} \\
\dd{\en_1 := {1 \over \pi} \int_{\supp \al}\! d\rho\; \rho^{d-3 \over 2}\;\gal(\rho)\,\l({3 \over 2}\,\rho - \gal(\rho)^2\r)\;.} \label{endef}
\end{array}
\end{equation}
Keeping in mind that $\al$ is assumed to have compact support, let us fix arbitrarily $\rho_1 \in (0,+\infty)$ such that
\begin{equation*}
\supp \al \subseteq [0,\rho_1] \;.
\end{equation*}
Next, let us consider the representation (\ref{evI}) of $I_\al(\vu)$ and notice that, for any $\vu > \rho_1$, it can be re-expressed as
\begin{equation*}
I_{\al}(\vu) := {\vu^{-3/2} \over \pi} \int_{0}^{\rho_1} d\rho\; {\rho^{d-3 \over 2}\,\gal(\rho) \over \sqrt{1\!-\!{\rho \over \vu}}\, \l(1 + {\gal(\rho)^2- \rho \over \vu}\r)} \;. \label{evInfA}
\end{equation*}
To proceed, notice that for any $N \in \{0,1,2,...\}$ and any fixed $\rho \in (0,\rho_1)$ there exits a family of coefficients $(\dn_n(\rho))_{n = 0,...,N}$ and a Taylor-Lagrange reminder function $T_N(\rho; \cdot) : (\rho_1,+\infty) \to \R$, $\vu \mapsto T_N(\rho;\vu)$, such that
\begin{equation*}
{\gal(\rho) \over \sqrt{1\!-\!{\rho \over \vu}}\l(1\!+\!{\gal(\rho)^2\!-\! \rho \over \vu}\r)} = \sum_{n = 0}^N \dn_n(\rho)\,\vu^{-n} + T_N(\rho,\vu) \quad \mbox{for all $\vu > \rho_1$}\;.
\end{equation*}
Let us mention that the coefficients $\dn_n(\rho)$ ($n = 0,...,N$) are all determined by integer powers of $\al(\rho)$; for example, one has
\begin{equation*}
\dn_0(\rho) := \gal(\rho)\;, \qquad \dn_1(\rho) := \gal(\rho) \l({3 \over 2}\,\rho - \gal(\rho)^2\r), \qquad \mbox{...}
\end{equation*}
Therefore, since $\al$ is assumed to be bounded, one easily infers that all the functions $\rho \mapsto \dn_n(\rho)$ ($n = 0,...,N$) are also uniformly bounded on $(0,\rho_1)$. \\
Concerning the reminder $T_N$, one has
\begin{equation*}
|T_N(\rho;\vu)| \leq S_{N+1}(\rho)\; \vu^{-(N+1)} \qquad \mbox{for all $\rho \in (0,\rho_1)$, $\vu \in (\rho_1+1,+\infty)$} \;, \label{asymInf}
\end{equation*}
where we have introduced the positive-valued function
\begin{equation*}
S_{N+1}(\rho) := \sup_{\vu \in (\rho_1+1,+\infty)} \l|{1 \over (N+1)!}{d^{N+1} \over d\vu^{N+1}}\l({\gal(\rho) \over \sqrt{1\!-\!{\rho \over \vu}}\l(1\!+\!{\gal(\rho)^2\!-\! \rho \over \vu}\r)}\r)\r|\;;
\end{equation*}
the latter can be easily proved to be uniformly bounded on $(0,\rho_1)$. \\
Summing up, the above results allow to infer that
\begin{equation*}
I_{\al}(\vu) = {\vu^{-3/2} \over \pi}\! \sum_{n = 0}^N \!\l(\int_{0}^{\vu_0}\!\!\! d\rho\; \rho^{d-3 \over 2}\dn_n(\rho)\r)\!\vu^{-n} + O(\vu^{-N-{5 \over 2}}) \quad \mbox{for $\vu \to +\infty$}\;, \label{emvIas1}
\end{equation*}
thus proving Eq.s \eqref{evIinf} - \eqref{endef}. \\

The above results on $I_\al(\vu)$, along with the explicit expression (\ref{evExp}) of the relative spectral measure, allow to infer straightforwardly the following facts concerning the map $(0,+\infty) \to \R$, $v \mapsto \eu(v)$: \\
i) $\eu \in C^0((0,+\infty);\R)$; in particular, $\eu$ is locally bounded on $(0,+\infty)$. \\
ii) There holds
\begin{equation}
\eu(v) = O(v^{d-2}) \qquad \mbox{for $v \to 0^+$}. \label{ev0}
\end{equation}
iii) For any $N \in \{0,1,2,...\}$, there holds the asymptotic expansion
\begin{equation}
\eu(v) = {1 \over v^2} \sum_{n = 0}^N \tilde{\en}_n\;v^{-2n} + O(v^{-2-2(N+1)}) \qquad \mbox{for $v \to +\infty$}\;, \label{evinf}
\end{equation}
where the real coefficients $(\tilde{\en}_n)_{n = 0,1,2,...}$ are related to the coefficients $(\en_n)_{n = 0,1,2,...}$ appearing in Eq. (\ref{evIinf}) by the rescaling
\begin{equation}
\tilde{\en}_n := {\pi \over (2\pi)^{{d + 1 \over 2}} \Gamma({d-1 \over 2})}\;\en_n \;. \label{pn0}
\end{equation}

\subsection{Relative zeta function in terms of the relative spectral measure}
\label{s:zeta1}

This section is devoted to the proof of identity \eqref{zetaint}. \\
The function $\zeu(s)$ is defined in terms of the integral kernel of the operator $Q^{rel} =e^{-tA_\alpha} - e^{-tA_0}$ by Eq. \eqref{relzeta1}. We start by recalling the well known identity 
\begin{equation}
\label{time}
\Q^{rel}(t;x^1,y^1,\xp-\yp)= -\frac{1}{2\pi i} \int_{\Lambda_\ve}dz\;e^{-zt} \RRK^{rel}(z;x^1,y^1,\xp-\yp) \;,
\end{equation}
where, for any $\ve>0$, $\Lambda_\ve$ is the contour 
\[\Lambda_\ve = \CC_\ve \cup \Lambda^+_\ve \cup \Lambda^-_\ve\qquad \mbox{where} \]
\[\CC = \{z = \ve e^{i\theta}| \;\theta\in[\pi/2,3\pi/2]\}\;, \qquad \Lambda_\ve^\pm = \{z = \lambda\pm i\ve| \; \lambda\in [0,+\infty)\},\]
and the integral over $\Lambda_\ve$ is taken counterclockwise. From the definition of $\zeu(s)$, see Eq. \eqref{relzeta1}, and identity \eqref{time}  one has  
\begin{equation*}\label{forever}
\zeu(s) =- \frac{1}{2\pi i \Gamma(s)} \int_0^{+\infty} dt \,t^{s-1} \int_{\R} dx^1 \int_{\Lambda_\ve}dz\;e^{-zt}\RRK^{rel}(z;x^1,x^1,\oo).
\end{equation*}
We claim that for all $\ve>0$ and $t>0$ 
\begin{equation*}\label{forever2}
\int_{\R} dx^1 \int_{\Lambda_\ve}dz\;e^{-zt}\RRK^{rel}(z;x^1,x^1,\oo) =\int_{\Lambda_\ve}dz\; e^{-zt}\ru(z)\;.
\end{equation*}
To prove the latter identity it is enough to show that the integrals can be exchanged. To this aim we note that, by Eq. \eqref{toolong} and our assumptions on $\alpha$ one has  
\begin{equation}
\label{all}
\begin{split}
\int_\R dx^1 \int_{\Lambda_\ve}   dz 
\int_{\R^{d-1}}\! d\kp\;    \frac{|e^{-zt}|\, \al(|\kp|^2) e^{-2|x^1|\Im \sqrt{z-|\kp|^2}}}{\left|\sqrt{z - |\kp|^2} \right| \left| \sqrt{z-|\kp|^2} +  i\al(|\kp|^2)/2 \right|} \\ 
\leq \|\alpha\|_\infty
 \int_{\Lambda_\ve} dz \int_{\{|\kp|^2 \in \supp\al\}}\! d\kp\; \frac{|e^{-z t}|}{\left(\Im \sqrt{ z - |\kp|^2}\right)^3 } ;
\end{split}\end{equation}
moreover, using the inequality
\[\frac1{\Im\sqrt{z-|\kp|^2}}\leq\frac{C}{\sqrt\ve} \max\left\{1,\sqrt\frac{\lambda}{\ve}, \frac{|\kp|}{\sqrt\ve}\right\} \qquad \text{for all } z\in\Lambda_\ve, \] 
one can easily prove that the integral on the r.h.s. of Eq. \eqref{all} is bounded for all $\ve>0$ and $t>0$. This suffices to infer that the integrals can be exchanged, as stated previously.\\ 
Next we note that, by the analyticity properties of $\ru(z)$, there exists a sequence $\ve_n\to 0^+$ such that 
\begin{align}
\zeu(s) & =- \frac{1}{2\pi i \Gamma(s)} \int_0^{+\infty} dt \,t^{s-1}\int_{\Lambda_\ve}dz\; e^{-zt}\ru(z) \nonumber \\ 
& = - \frac{1}{2\pi i \Gamma(s)} \int_0^{+\infty} dt \,t^{s-1} \lim_{n\to+\infty}\int_{\Lambda_{\ve_n}}dz\; e^{-zt}\ru(z).
\label{her} 
\end{align}
We claim that for all $t>0$ one has 
\begin{equation}
\label{patchwork}
\frac{1}{2\pi i} \lim_{\ve\to0^+}\int_{\Lambda_\ve}dz\;e^{-zt} \ru(z) =- \int_{0}^{+\infty} dv\,e^{-v^2t} \eu(v ).
\end{equation}
Then the  identity  \eqref{zetaint}  follows from Eq.s \eqref{her} and \eqref{patchwork}, by exchanging order of integration and by taking into account the identity $\frac{1}{\Gamma(s)} \int_0^\infty dt \, t^{s-1} e^{-v^2 t} = v^{-2s}$. \\
To prove Eq. \eqref{patchwork} we start by noticing that for all $t>0$
\begin{equation}
 \int_{\Lambda_\ve}dz\;e^{-zt} \ru(z) 
 =-  \lim_{\ve\to 0^+}  \int_{0}^{+\infty} d\lambda \left(e^{-(\lambda+i\ve)t} \ru(\lambda+i\ve) - e^{-(\lambda-i\ve)t} \ru(\lambda-i\ve) \right) , \label{rama}
\end{equation}
where we used the fact that 
\[\lim_{\ve\to0^+} \int_{\CC_\ve}dz\; e^{-zt} \ru(z) = 0\;.\]
The latter claim follows from the bound 
\[|\ru(\ve e^{i\theta})| \leq C \frac{\|\al\|_\infty }{\delta} \int_{\{|\kp|^2 \in \supp\al\}}\! d\kp\; {1\over  \sqrt{\ve^2+ |\kp|^4}} \qquad \forall z\in \CC_\ve, \vspace{-0.2cm} \] 
where we used Eq. \eqref{rzkp} and the fact that $|z-|\kp|^2|^{-1} \leq(\ve^2+|\kp|^4)^{-1/2}$ for all $z\in\CC_\ve$.
In fact, in view of the estimate $\int_{\{|\kp|^2 \in \supp\al\}}\! d\kp\; (\ve^2+ |\kp|^4)^{-1/2} \leq C\, \ve^{-3/4}$ for all $0<\ve<1$ (where $C$ is a constant that depends on $d$ and $\supp\al$), the mentioned bound allows to infer $\l|\int_{\CC_\ve}dz\; e^{-zt} \ru(z)\r| \leq C\,\eps^{1/4}$, thus proving the previous claim. \\
To move the limit inside the integral in Eq. \eqref{rama} we use dominated convergence theorem. To this aim, we use first the trivial bound 
\begin{equation}\label{glass}
\begin{split}
\left|e^{-(\lambda+i\ve)t} \ru(\lambda+i\ve) - e^{-(\lambda-i\ve)t} \ru(\lambda-i\ve) \right|   \\ 
 \leq 
e^{-\lambda t} \left|2\sin (\ve t) \ru(\lambda+i\ve) \right| 
+e^{-\lambda t} \left|\ru(\lambda+i\ve) - \ru(\lambda-i\ve) \right| .
\end{split}
\end{equation}
Next we note that 
\[e^{-\lambda t} \left|2\sin (\ve t) \ru(\lambda+i\ve) \right|  \leq C e^{-\lambda t} t  \sqrt{\ve}  \frac{\|\alpha\|_\infty}{\delta} \int_{\{|\kp|^2 \in \supp \alpha\}} d\kp \frac1{\left|\lambda -|\kp|^2\right|^{1/2}}.  \]
Hence the limit for $\ve\to 0^+$ of the first term at the r.h.s. of Eq.\eqref{glass} is zero. On the other hand, for the second term at the r.h.s. we use, see Eq. \eqref{rzkp}, 
\[\begin{split}
|\ru(\lambda+i\ve) - \ru(\lambda-i\ve)|
\leq C\|\alpha\|_\infty  \Bigg( \frac1\delta\int_{\{|\kp|^2 \in \supp\al\}}\! d\kp\; \frac{\ve}{(\lambda- |\kp|^2)^2+\ve^2}  \\ 
+\frac{1}{\delta^2} \int_{\{|\kp|^2 \in \supp\al\}}\! d\kp\; \frac{\big|\sqrt{\lambda -|\kp|^2+i\ve} -\sqrt{\lambda -|\kp|^2-i\ve}\big|}{\big|\lambda  - |\kp|^2+i\ve\big|} \Bigg)\,.
\end{split}\]
The first term is uniformly bounded in $\ve$. For the second term we use the inequality 
\[\frac{\big|\sqrt{\rho+i\ve} -\sqrt{\rho-i\ve}\big|}{\big|\rho+i\ve\big|} \leq \frac{C}{\sqrt{|\rho|}},\]
 which holds true for all $\rho\in \R$ and $\ve>0$, from which it follows that the second term is uniformly bounded as well. Identity \eqref{patchwork} follows  from the dominated convergence theorem applied to Eq. \eqref{rama} together with the  change of variables $\lambda \to v^2$. 

\subsection{Analytic continuation of the relative zeta function}\label{SecZeta}\vspace{-0.2cm}
In this section we obtain the analytic continuation of the relative zeta function $\zeu(s)$. We start with the representation \eqref{zetaint}. In view of the continuity of the map $v \mapsto e(v)$ on $(0,+\infty)$ and of its asymptotic behaviours for $v \to 0^+$ and $v \to +\infty$ discussed previously (see, in particular, Eq.s (\ref{ev0}) and (\ref{evinf})), it appears that the integral in Eq. (\ref{zetaint}) converges for any complex $s$ in the strip given in Eq. \eqref{Res1}.\\
To proceed, let us recall the well-known fact that the asymptotic expansions of $e(v)$ for $v \to 0^+$ and $v \to +\infty$ can be used to construct explicitly the analytic continuation of the map $s \mapsto \zeta(s)$ to larger regions of the complex plane, by means of standard techniques. Let us point out that, for the applications to be discussed in the forthcoming Section \ref{s:SecCas}, it suffices to determine the said analytic continuation of $\zeta(s)$ to regions with larger negative values of $\Re s$
(in particular, to a neighbour of $s = -1/2$); to this purpose, let us fix $v_0 \in (0,+\infty)$ arbitrarily and re-express the relative partial zeta function as
\begin{equation}
\begin{array}{c}
\dd{\zeta(s) = \zeta^{(<)}(s) + \zeta^{(>)}(s)\;,} \vspace{0.1cm} \\
\dd{\zeta^{(<)}(s) := \int_0^{v_0} dv\;v^{-2s}\;e(v)\;, \qquad \zeta^{(>)}(s) := \int_{v_0}^{+\infty} dv\;v^{-2s}\;e(v)\;.}
\end{array} \label{zetaintMm}
\end{equation}
Notice that the asymptotic behaviour in Eq. (\ref{ev0}) suffices to infer that the map $s \mapsto \zeta^{(<)}(s)$ is analytic for $\Re s < (d-1)/2$. On the other hand, the integral defining $\zeta^{(>)}(s)$ only converges for $\Re s > -1/2$; in order to construct its analytic continuation to larger negative $\Re s$, let us fix $N \in \{0,1,2,...\}$ and proceed to add and subtract to the integrand in Eq. (\ref{zetaintMm}) the first $N+1$ terms of the asymptotic expansion of $e(v)$ for $v \to +\infty$ (see Eq. (\ref{evinf})\,); we thus obtain
\begin{equation*}
\zeta^{(>)}(s) = \sum_{n = 0}^N \tilde{\en}_n \int_{v_0}^{+\infty}\!\! dv\;v^{-2s-2 - 2n} + \int_{v_0}^{+\infty}\!\! dv\;v^{-2s}\l(e(v) - {1 \over v^2} \sum_{n = 0}^N \tilde{\en}_n\;v^{-2n}\r) . \label{zetaintAC0}
\end{equation*}
Therefore, using the elementary identity
\begin{equation*}
\int_{v_0}^{+\infty}\!\!\! dv\;v^{-2s-2 - 2n} = {v_0^{-2s-2n-1} \over 2s+2n+1} \quad \mbox{for all $s\!\in\!\C$, $n\!\in \!\N$ with $\Re s > - n - {1 \over 2}$}\;,
\end{equation*}
one obtains
\begin{equation}
\zeta^{(>)}(s) = \sum_{n = 0}^N {v_0^{-2s-2n-1} \over 2s+2n+1}\; \tilde{\en}_n + \int_{v_0}^{+\infty}\!\! dv\;v^{-2s}\l(e(v) - {1 \over v^2} \sum_{n = 0}^N \tilde{\en}_n\;v^{-2n}\r) . \label{zetaintAC}
\end{equation}
Even though the above expression was derived under the assumption $\Re s > -1/2$, the following facts are apparent. On the one hand, the first term on the right-hand side of Eq. (\ref{zetaintAC}) is a sum of functions which are meromorphic on the whole complex plane, with only simple poles at the points $\{-1/2,-3/2,...,-N -1/2\}$; on the other hand, the second term in Eq. (\ref{zetaintAC}) is an integral which converges by construction for any $\Re s > - N - 3/2$ and defines an analytic function of $s$ in this region. \\
Summing up, the above arguments allows to infer that the identity
\begin{equation}
\begin{array}{c}
\dd{\zeta(s) = } \\
\dd{\sum_{n = 0}^N {v_0^{-2s-2n-1} \over 2s+2n+1}\; \tilde{\en}_n + \int_0^{v_0}\!\! dv\;v^{-2s}\;e(v) + \int_{v_0}^{+\infty}\!\!\! dv\;v^{-2s}\!\l(\!e(v) - {1 \over v^2} \sum_{n = 0}^N \tilde{\en}_n\,v^{-2n}\!\r)}
\end{array} \label{zetaintAC1_1}
\end{equation}
determines the analytic continuation of the map $s \mapsto \zeta(s)$ to a function which is meromorphic in the larger strip
\begin{equation}\label{zetaintAC1_2}
\l\{ s \in \C \,\;\Big|\; - {3 \over 2} - N < \Re s < {d-1\over 2} \r\}\;, 
\end{equation}
with possible simple pole singularities at the points
\begin{equation}
\label{zetaintAC1_3}
\{-1/2,-3/2,...,-N- 1/2\}\;.
\end{equation}

\section{The thermal Casimir energy}
\label{s:SecCas}
We work in natural units, meaning that we fix the speed of light $c$, the reduced Plank constant $\hbar$ and the Boltzmann constant $\kappa$ as follows:
\begin{equation*}
c := 1 \;, \qquad \hbar := 1\;, \qquad \kappa := 1\;.
\end{equation*}
In this section we proceed to compute the renormalized Casimir energy per unit surface $\Ec(\beta)$ for a massless scalar field at temperature $T = 2\pi/\beta$ ($\beta \in (0,+\infty)$), living in $(d+1)$-dimensional spacetime. A simple adaptation of the arguments presented in \cite{spreafico-zerbini09} allows to infer that this observable is completely determined by the singular and regular parts of the Laurent expansion at $s = -1/2$ of the  relative zeta function $\zeu(s)$, discussed in the previous section; more precisely, there holds
\begin{equation}
\Ec(\beta) = {1 \over 2}\,\mbox{Res}_0\Big|_{s = -1/2}\! \zeu(s)\, +(1-\log (2\ell))\;\mbox{Res}_1\Big|_{s = -1/2} \zeu(s) + \partial_\beta\log \eta(\beta)\;, \label{EnC}
\end{equation}
where $\ell \in (0,+\infty)$ is a length parameter required by dimensional arguments and
\begin{equation}
\log \eta(\beta) := \int_0^{+\infty} dv\; \log(1-e^{-\beta v})\; \eu(v)\;. \label{logeta}
\end{equation}
In view of Eq. (\ref{zetaintAC1_1}), here employed with $ N = 0$ and any fixed $v_0 \in (0,+\infty)$, one readily infers the following
({\footnote{Following \cite{spreafico-zerbini09}, for any $s_0 \in \C$ we use the notation
$$ \mbox{Res}_n\Big|_{s = s_0} \zeu(s) := \l\{
\begin{array}{c}
\dd{\mbox{coefficient of $(s - s_0)^{-n}$}} \vspace{0.1cm} \\ \dd{\mbox{in the Laurent expansion of $\zeu(s)$ at $s = s_0$}}
\end{array}\r\}\;. $$}}):
\begin{equation}
\mbox{Res}_1\Big|_{s = -1/2}\zeu(s) = {1 \over 2}\;\tilde{\en}_0 \;, \label{Re1}
\end{equation}
\begin{equation}
\begin{array}{c}
\dd{ \mbox{Res}_0\Big|_{s = -1/2} \zeu(s) = - \tilde{\en}_0 \,\log v_0 + \int_0^{v_0}\!\! dv\;v\;\eu(v) + \int_{v_0}^{+\infty}\!\!\! dv\;v \l(\!\eu(v) - {\tilde{\en}_0 \over v^2} \!\r). }
\end{array} \label{Re0}
\end{equation}
Summing up, Eq.s (\ref{EnC}) - (\ref{Re0}) give the explicit expression for the renormalized Casimir energy
\begin{equation}
\Ec(\beta) = \label{EnCExp} \vspace{-0.1cm}
\end{equation}
$$ {1 \over 2}\l[\Big(\!1-\log (2\ell\,v_0)\!\Big)\tilde{\en}_0 + \!\int_0^{v_0}\!\! dv\;v\;\eu(v) + \!\int_{v_0}^{+\infty}\!\!\! dv\;v \l(\!\eu(v) - {\tilde{\en}_0 \over v^2} \!\r) + \!\int_0^{+\infty}\!\!\! dv\; {2\,v\;\eu(v) \over e^{\beta v} - 1}\r], $$
where the relative spectral measure $\eu(v)$ and the coefficient $\tilde{\en}_0$ are given, respectively, by Eq.s (\ref{evExp}) - (\ref{evI}) and Eq.s (\ref{endef}) and  (\ref{pn0}).\\
Before proceeding, let us remark that the first three terms in Eq. (\ref{EnCExp}) correspond to the zero temperature ($\beta \to +\infty$) contribution, while the last term gives the temperature correction.

\subsection{The Casimir energy for a simple model in spatial dimension $d = 3$}\label{subsec1}
As a simple application of the results derived previously, let us consider the $3$-dimensional configuration corresponding to the choice
\begin{equation*}
\al(\rho) = \al_0 \;\chi_{(0,K^2)}(\rho)\quad \mbox{for some $\al_0, K > 0$} \qquad (d = 3)\;; \label{aConst}
\end{equation*}
this clearly fulfils our assumptions on $\alpha$. In the following, in agreement with Eq. (\ref{galdef}), we put
\begin{equation}
\gal_0 := \al_0 /2\;. \label{gal0def}
\end{equation}
In this case, one can derive a simple, fully explicit expression for the corresponding thermal Casimir energy. To this purpose, let us first notice that the expression (\ref{evI}) of the integral $I_\al$ can be evaluated to give
\begin{equation*}
I_{\al}(\vu) = {2 \over \pi}\l[\arctan\l({\sqrt{\vu} \over \gal_0}\r) - \chi_{(K^2,+\infty)}(\vu)\,\arctan\l({\sqrt{\vu-K^2} \over \al_0}\,\r)\r] . \label{evIEs}
\end{equation*}
Substituting the above result in Eq. (\ref{evExp}) and making a few elementary manipulations,
one obtains for the relative spectral measure 
\begin{equation*}
\eu(v) = - {v \over 2\pi^2} \Bigg[\arctan\l({\gal_0 \over v}\r) - \arctan\l(\!{\gal_0 \over \sqrt{v^2-K^2}}\r)\chi_{(K,+\infty)}(v) \Bigg]\,, \label{evExp3}
\end{equation*}
On the other hand, Eq.s (\ref{endef}) and (\ref{pn0}) give straightforwardly
\begin{equation*}
\tilde{\en}_0 = {1 \over 4\pi^2}\,\gal_0\;K^2 \;.
\end{equation*}
Due to the above results and upon evaluation of some elementary integrals, for any $\beta,\ell \in (0,+\infty)$ and some fixed $v_0 \in (0,K)$ chosen arbitrarily, Eq. (\ref{EnCExp}) yields
\begin{equation*}
\begin{array}{c}
\dd{\Ec(\beta) = - {1 \over 12\pi^2} \Bigg[{\pi \over 2}\,K^3 + {3 \over 2}\,\gal_0\,K^2\l(\!\log(\ell\,R)\! - \!{7 \over 6} \r) + \,\gal_0^3\,\log\l(\!{2\gal_0 \over K}\!\r) - \,\Ecf(\gal_0,K)\; + }\vspace{0.05cm} \\
\dd{\hspace{2.5cm} + \int_0^{+\infty}\!\! dv\;\Bigg({6\,v^2 \over e^{\beta v} - 1} - {6\,v\,\sqrt{v^2+K^2} \over e^{\beta \sqrt{v^2+K^2}} - 1}\Bigg) \;\arctan\l(\!{\gal_0 \over v}\!\r)\Bigg] \;, }\\
\end{array} \label{EcRen}
\end{equation*}
where we have introduced the (continuous) function 
\begin{equation*}
\Ecf(x,y) := \l\{\begin{array}{ll} \dd{(x^2 - y^2)^{3/2}\; \mbox{arccoth}\Bigg(\!{x \over \sqrt{x^2 - y^2}}\Bigg)} & \quad \dd{\mbox{for $x \geq y$}\;,} \\
\dd{(y^2- x^2)^{3/2} \Bigg[{3\pi \over 2} - \arctan\Bigg(\!{x \over \sqrt{y^2-x^2}}\Bigg)\Bigg]} & \quad \dd{\mbox{for $x < y$}\;.}
\end{array} \r.
\end{equation*}
Note that the above result allows to infer by simple arguments that, in the limiting case where $\al_0$ is kept fixed and $K \to +\infty$, there holds
\begin{equation}
\begin{array}{c}
\dd{\Ec(\beta) = {\gal_0^3 \over 12\pi^2} \Bigg[\pi\,{K^3 \over \gal_0^3} - {3 \over 2}\,{K^2 \over \gal_0^2}\l(\! \log(\ell\,K)\! - \!{1 \over 2}\r) - {9 \pi \over 4}\,{K \over \gal_0} + \log(\ell K) \; + } \vspace{0.05cm} \\
\dd{\hspace{0.7cm} + \l({4 \over 3} - \log(2 \ell \gal_0)\r) - {1 \over \gal_0^3} \int_0^{+\infty}\!\! dv\;{6\,v^2 \over e^{\beta v} - 1}\;\arctan\l(\!{\gal_0 \over v}\!\r)\! +\; O\Big({\gal_0 \over K}\Big)\Bigg]\;. }\\
\end{array} \label{EcAsym}
\end{equation}
Obviously enough, the above result allows to make a comparison with the model corresponding to a constant, not compactly supported function 
\begin{equation}
\al(\rho) = \al_0 \quad \mbox{for all $\rho \in [0,+\infty)$ and some $\al_0 > 0$} \qquad (d = 3)\;, \label{aConst2}
\end{equation}
that is the model typically considered in the literature \cite{Bord, Graham, Khus, MaTru, Cast}; as reviewed in Appendix A, in this case the Casimir energy is given by
\begin{equation}
\begin{array}{c}
\dd{\Ec(\beta) = {\gal_0^3 \over 12\pi^2}\l({4 \over 3} - \log(2 \ell \gal_0)\r) - {1 \over 2\pi^2} \int_0^{+\infty}\!\! dv\;{v^2 \over e^{\beta v} - 1}\;\arctan\l(\!{\gal_0 \over v}\!\r) \;. }
\end{array} \label{EnC3}
\end{equation}
This appears to coincide with the finite, ``renormalized'' part of the asymptotic expansion (\ref{EcAsym}), which is obtained removing by brute force the divergent terms in the cited expansion.

\appendix
\section*{Appendix A. The case $\al = const.$ in spatial dimensions $d = 3$}\label{subd23}
In order to make connection with the existing literature, in the present appendix we briefly review the computation of the Casimir energy for the $3$-dimensional model described in Eq. (\ref{aConst2}), that is
$$ \al(\rho) = \al_0 \quad \mbox{for all $\rho \in [0,+\infty)$ and some $\al_0 > 0$} \qquad (d = 3)\;. $$
As a matter of fact, it can be easily checked that all the arguments described in the present manuscript continue to make sense also in this particular case, even though $\al$ does not fulfil the required assumptions  since it does not have compact support. \\
First of all, let us notice that the integral representation (\ref{rz}) for the function $\ru(z)$ continues to make sense for any $z \in \C \setminus [0,+\infty)$ (the integral in the cited equation is trivially seen to remain finite in the present case). Then, by the same arguments of Section \ref{s:ev} one obtains an expression like Eq. (\ref{evExp}) for the relative spectral measure $\eu(v)$; moreover, the term $I_\al$ appearing therein (given by Eq. (\ref{evI})\,) can be evaluated explicitly. This allows to infer for the relative spectral measure the expression
\begin{equation}
\eu(v) = - {v \over 2\pi^2}\;\arctan\l({\gal_0 \over v}\r)\,\chi_{(0,+\infty)}(v)\;, \label{evExpC1} 
\end{equation}
where $\gal_0$ is defined according to Eq. (\ref{gal0def}). This shows that the map $v \mapsto \eu(v)$ is continuous on $(0,+\infty)$ and fulfils, for any $N \in \{0,1,2,...\}$,
\begin{equation*}
\eu(v) = \l\{ \begin{array}{ll} \dd{O(v)} & \quad \dd{\mbox{for $v \to 0^+$}}\;, \\
\dd{\sum_{n = 0}^N {(-1)^{n+1}\,\gal_0^{2n+1} \over 2\pi^2 (2n + 1)}\;v^{-2n} + O(v^{-2(N+1)})} & \quad \dd{\mbox{for $v \to +\infty$}\;.} \\
\end{array}\r.
\end{equation*}
Next, let us consider the representation (\ref{zetaint}) of the relative zeta function $\zeu(s)$ in terms of $\eu(v)$; in view of the above considerations, it appears that the integral in the cited equation is finite for any complex $s$ inside the strip 
\begin{equation*}
\Big\{ s \in \C \;\Big|\; {1 \over 2} < \Re s < 1 \Big\}\;.
\end{equation*}
To proceed, notice that for any such $s$ the integral in Eq. (\ref{zetaint}) can be evaluated explicitly using the expression (\ref{evExpC1}) for $\eu(v)$; one obtains
\begin{equation*}
\zeu(s) = - {\gal_0^{2-2s} \over 8\pi\,(s-1)\,\cos(\pi s)}\;, \label{expZeta}
\end{equation*}
which determines the analytic continuation of $s \mapsto \zeu(s)$ to a function which is meromorphic on the whole complex plane, with simple pole singularities at $s = 1$ and $s = \pm 1/2, \pm 3/2, ...$\,. \\
Using the above expression for $\zeu(s)$ and Eq.s (\ref{EnC})-(\ref{logeta}) for the thermal Casimir energy $\Ec(\beta)$, one easily infers the final result (\ref{EnC3}), that is
$$ \begin{array}{c}
\dd{\Ec(\beta) = {\gal_0^3 \over 12\pi^2} \l({4 \over 3} - \log (2\ell \gal_0)\!\r) - {1 \over 2\pi^2} \int_0^{+\infty} dv\; {v^2 \over e^{\beta v} - 1} \;\arctan\l({\gal_0 \over v}\r)\,. }\\
\end{array} $$
For completeness, let us mention that the above expression can be easily employed to derive the zero temperature expansion ($\beta \to +\infty$) of the Casimir energy $\Ec(\beta)$.

\bigskip
{\bf Acknowledgments.} C.C. and D.F. acknowledge the support of the FIR 2013 project ``Condensed Matter in Mathematical Physics'', Ministry of University and
Research of Italian Republic  (code RBFR13WAET).

\end{document}